\documentclass{article}

\usepackage{me}

\evenmargins{1in}

\begin{document}

\def\emb#1{{\it#1\/}}
\def\sb#1{{\sl#1\/}}
\def\-#1{\overline{#1}}
\def\(#1){^{(#1)}}

\def\O{\tilde O}

\def\C{{\cal C}}

\def\val(#1,#2){\hbox{\rm VAL}(#1,#2)}
\def\valv(#1){#1}
\def\w(#1,#2){\hbox{\rm wt}(#1,#2)}
\def\E[#1]{\hbox{\rm E}[#1]}
\def\c{\tilde k}
\def\const{\rho}
\def\gmul#1#2{#1 \times #2}
\def\gsamp#1#2{#2(#1)}
\def\gcomp#1#2{#2[#1]}

  \def\sqr#1#2{{\vcenter{\vbox{\hrule height .#2pt \hbox{\vrule
width
.#2pt height#1pt \kern#1pt \vrule width .#2pt} \hrule height
.#2pt}}}}


\title{Randomized Approximation Schemes for Cuts and Flows in Capacitated Graphs}

\author{Andr\'as A.\ Bencz\'ur\thanks{Supported from grants 
         OTKA T-30132 and T-29772; NWO-OTKA; AKP 104024 \hfill\break
         E-mail: \texttt{benczur@sztaki.hu} \hfill\break
         URL: \texttt{http://www.sztaki.hu/\~{ }benczur}}\\       
	 Computer and Automation Institute\\ 
	 Hungarian Academy of Sciences, and\\ 
         Department of Operations Research\\ 
	 E\"otv\"os University, Budapest
\and David R. Karger\thanks{M.I.T.  Laboratory for Computer Science,
         545 Technology Square, Cambridge MA 02139. E-mail: {\tt
         karger@theory.lcs.mit.edu}. 
  \protect\newline URL: {\tt http://theory.lcs.mit.edu/\~\/karger}.
  Research supported in part by NSF award CCR-9820978 and a Packard
         Foundation Fellowship.}\\ 
Laboratory for Computer Science\\M.\ I.\ T.} 
\maketitle


\begin{abstract}
  We improve on random sampling techniques for approximately solving
  problems that involve cuts and flows in graphs.  We give a
  near-linear-time construction that transforms any graph on $n$
  vertices into an $O(n\log n)$-edge graph on the same vertices whose
  cuts have approximately the same value as the original graph's.  In
  this new graph, for example, we can run the ${\tilde
  O}(m^{3/2})$-time maximum flow algorithm of Goldberg and Rao to find
  an $s$--$t$ minimum cut in ${\tilde O}(n^{3/2})$ time.  This
  corresponds to a $(1+\epsilon)$-times minimum $s$--$t$ cut in the
  original graph.  In a similar way, we can approximate a sparsest cut
  to within $O(\log n)$ in $\O(n^2)$ time using a previous
  $\Olog(mn)$-time algorithm.  A related approach leads to a
  randomized divide and conquer algorithm producing an approximately
  maximum flow in $\O(m\sqrt{n})$ time.
\end{abstract}

\section{Introduction}

Previous work~\cite{Karger:Thesis,Karger:Skeleton,Karger:Lincut} has
shown that random sampling is an effective tool for problems involving
cuts in graphs.  A {\em cut} is a partition of a graph's vertices into
two groups; its {\em value} is the number, or in weighted graphs the
total weight, of edges with one endpoint in each side of the cut.
Many problems depend only on cut values. The maximum flow that can be
routed from $s$ to $t$ is the minimum value of any cut separating $s$
and $t$~\cite{Ford:Maxflow}. A minimum bisection is the smallest cut
that splits the graph into two equal-sized pieces.  The
\emph{connectivity} or \emph{minimum cut} of the graph, which we
denote throught by $c$, is equal to the minimum value of any cut.

Random sampling ``preserves'' the values of cuts in a graph.  If we
pick each edge of a graph $G$ with probability $p$, we get a new graph
in which every cut has expected value exactly $p$ times it value in
$G$.  A theorem by Karger~\cite{Karger:Skeleton} shows that if the
graph has unit-weight edges and minimum cut $c$, then sampling with
probability roughly $1/\epsilon^2 c$ gives cuts that are all, with
high probability, within $1\pm\epsilon$ of their expected values.  In
particular, the minimum cut of the sampled graph corresponds to a
$(1+\epsilon)$-times minimum cut of the original graph.  Similarly, an
$s$-$t$ minimum cut of the sampled graph is a $(1+\epsilon)$-times
minimum $s$-$t$ cut of the original graph.  Since the sampled graph
has fewer edges (by a factor of $1/c$ for any fixed $\epsilon$),
minimum cuts can be found in it faster than in the original graph.
Working through the details shows that an approximately minimum cut
can be found roughly $c^2$ times faster than an exact solution.

A variant of this approach finds approximate solutions to flow
problems via \emph{randomized divide and conquer}.  If we randomly
partition the edges of a graph into roughly $\epsilon^2 c$ subsets,
each looks like the sample discussed in the previous paragraph, so has
approximately accurate cuts.  In other words, random division is a
good approximation to evenly dividing up the capacities of all the
cuts.  By max-flow min-cut duality~\cite{Ford:Maxflow}, this means tha
the $s$-$t$ max-flow of $G$ is also approximately evenly divided up.
We can find a maximum flow in each of the subgraphs and add them
together to get a flow in $G$ that is at least $(1-\epsilon)$ times
optimal.  Again, detailed analysis shows that finding this approximate
flow can be done $c$ times faster than finding the exact maximum flow.

Unfortunately, the requirement that $p=\Omega(1/c)$ limits the
effectiveness of this scheme.  For cut approximation, it means that in
a graph with $m$ edges, we can only reduce the number of edges to
$m/c$.  Similarly for flow approximation, it means we can only divide the
edges into $c$ groups.  Thus, when $c$ is small, we gain little.
Results can be even worse in weighted graphs, where the ratio of total
edge weight to minimum cut value is unbounded.

\subsection{Results}

In this paper, we show how {\em nonuniform sampling} can be used to
remove graph sampling's dependence on the minimum cut $c$.  Our main
results are 
twofold: one for cut problems, and one for flow problems.  For cuts,
we show that by sampling edges \emph{nonuniformly}, paying greater
attention to edges crossing small cuts, we can produce accurate
samples with far less than $m/c$ edges---rather, the resulting
\emph{compressed} graph has only $\O(n/\epsilon^2)$ edges, regardless
of the number of edges in the original graph.\footnote{The notation
$\Olog(f)$ denotes $O(f \polylog I)$ where $I$ is the input problem
size.}  In consequence, we show that a $(1+\epsilon)$-times minimum
$s$-$t$ cut can be found in $\O(n^{3/2}/\epsilon^3)$ time in general
capacity graphs (as compared to the $\O(m^{3/2})$ exact bound) and
$\O(nv/\epsilon^2)$ time in unit-capacity graphs with flow value $v$
(as compared with the $O(mv)$ exact bound).  Similarly, a nonuniform
divide-and-conquer approach can be used to find a $(1-\epsilon)$ times
maximum flow in $\O(m\sqrt{n}/\epsilon)$ time.  Our approach works for
undirected graphs with arbitrary weights (capacities).

Even ignoring the algorithmic aspects, the fact that any graph can be
approximated by a sparse graph is of independent combinatorial
interest.

In addition to proving that such sampling works, we give fast
algorithms for determining the importance of different edges and the
correct sampling probabilities for them.  This involves an extension
of the {\em sparse certificate} technique of Nagamochi and
Ibaraki~\cite{Nagamochi:Connectivity}.

Using these results, we demonstrate the following:

\begin{theorem}
  Given a graph $G$ and an error parameter $\epsilon$, there is a
graph $G'$ such that
\begin{itemize}
\item $G'$ has  $O(n \log n\,/\epsilon^2)$ edges and
\item the value of every cut in $G'$ is $(1\pm \epsilon)$ times the
  value of the corresponding cut in $G$.
\end{itemize}
$G'$ can be constructed in $O(m\log^2 n)$ time if $G$ is unweighted
and in $O(m\log^3 n)$ time if $G$ is weighted.
\end{theorem}

It follows that given any algorithm to (even approximately) solve a cut
problem, if we are willing to accept an approximate answer, we can
substitute $n \log n$ for any factor of $m$ in the running time.  Our
 applications of this result are the following:

\begin{corollary}
In an undirected graph, a $(1+\epsilon)$ times minimum $s$--$t$ cut
can be found in $\O(n^2/\epsilon^2)$ or $\O(n^{3/2}/\epsilon^3)$ time.
\end{corollary}

\begin{corollary}
In an undirected graph, a $(1+\epsilon)$ times minimum $s$--$t$ cut of
value $v$ can be found in $\O(nv/\epsilon^2)$ time.
\end{corollary}

\begin{corollary}
  An $O(\log n)$-approximation to the sparsest
  cut in an undirected graph can be found in $\O(n^2/\epsilon^2)$ time.
\end{corollary}

These corollaries follow by applying our sampling scheme to
(respectively) the maximum flow algorithms of Goldberg and
Tarjan~\cite{Goldberg:Maxflow} and Goldberg and
Rao~\cite{Goldberg:CapacitatedBlocking}, the classical
augmenting-paths algorithm for maximum
flow~\cite{Ford:Maxflow,Ahuja:Flow}, and the Klein-Stein-Tardos
algorithm for approximating the sparsest
cut~\cite{Klein:Concurrent}.

A related approach helps solve flow problems: we divide edges crossing
small cuts into several parallel pieces, so that no one edge forms a
substantial fraction of any cut it crosses.  We can then apply a
randomized divide and conquer scheme.  If we compute a maximum flow in
each of the subgraphs created by the random division using the
Goldberg-Rao algorithm, and then add the flows into a flow in $G$, we
deduce the following corollary:

\begin{corollary}
  A $(1-\epsilon)$ times maximum flow can be found in
  $\Olog(m\sqrt{n}/\epsilon)$ time.
\end{corollary}

The work presented here combines work presented earlier by Karger and
Benczur~\cite{Karger:Stcut} and by Karger~\cite{Karger:SmoothFlow}.
The presentation is simplified and slight improvements are given.

\subsection{Method}

The previous work on sampling for cuts is basically an application of
the Chernoff bound.  Our goal in cut sampling is to estimate the total
weight (or number, in the case of unit-weight graphs) of edges
crossing each cut of the graph.  We motivate our approach by
considering a simpler problem---that of estimating a single cut.
Consider a set of $m$ weights $w_e$, and suppose that we wish to
estimate the sum $S=\sum w_e$.  A natural approach is random sampling:
we choose a random subset of the weights, add them, and scale the
result appropriately.  A somewhat easier to analyze approach is to
choose each weight independently with some probability $p$, compute
their sum $S'$, and estimate $S=S/p$.  Since we choose only $pm$
weights in expectation, this sampling approach saves time.  But we
must analyze its accuracy.  The Chernoff bound is a natural tool.

\begin{lemma}[Chernoff~\cite{Chernoff}]
Given any set of random variables $X_i$ with values distributed in the
range $[0,1]$, let $\mu=E[\sum X_i]$ and let $\epsilon < 1$.  Then 
\[
\Pr[\sum X_i \notin (1\pm\epsilon)\mu] \le 2e^{-\epsilon^2\mu/3}.
\]
\end{lemma}

The lemma's requirement that $X_i \le 1$ is in force to prevent any
one random variable from ``dominating'' the outcome of the sampling
experiment.  For example, if one variable takes on value $S$ with
probability $1/S$ and 0 otherwise, while all other variables are
uniformly 0, then the (relatively rare, but still occasional) outcome
of taking on value $S$ will dramatically skew the sum away from is
expectation of $1$.

We can model our sampling experiment so as to apply the Chernoff
bound.  For now, let us assume that each $w_e \le 1$.  Let $X_e$ be a
random variable defined by setting $X_e=w_e$ with probability $p$ and
$X_e=0$ otherwise.  Note that $\sum X_e$ is the value of our sampling
experiment of adding the weights we have chosen to examine.  Also,
$E[\sum X_e]=\sum pw_e=pS$. The variables $X_e$ satisfy the conditions
of the Chernoff bound, letting us deduce that the probability that
$\sum X_e$ deviates by more than $\epsilon$ from its expectation is
$e^{-\epsilon^2 pS/3}$.  Note that this deviation is 
exponentially unlikely as a function of the expected sample value
$pS$.

We now note some slack in this sampling scheme.  If some $w_e \ll 1$,
then its random sample variable $X_e$, which takes on values 0 or
$w_e$, is far away from violating the requirement that each $X_e \in
[0,1]$.  We can afford to apply a more aggressive sampling strategy
without violating the Chernoff bound assumptions.  Namely, we we can
set $X_e=1$ with probability $p w_e$ and $0$ otherwise.  We have
chosen this probability because it keeps the expected value of each
$X_e$, and thus $E[\sum X_e]$, unchanged while making each variable
``tight'' against the $X_e\le 1$ limit of the Chernoff bound.  Since
this fits the preconditions of the lemma, we preserve the
$(1\pm\epsilon)$ concentration around the mean shown by the Chernoff
bound.  However, under this scheme, the expected number of sampled
values drops from $pm$ to $\sum pw_e$ (which is less since we assume
each $w_e < 1$).  This is a noteworthy quantity: it is equal to the
expected value $\mu=E[\sum X_e]$.  Since the probability of error in
the Chernoff bound is itself a function only of $\mu$, it follows that
under this scheme the expected number of samples $\mu$ needed to
guarantee a certain error probability $\delta$ is a function only of
the desired bound (namely, $\mu=3(\ln 1/\delta)/\epsilon^2$), and not
of the number of variables $m$ or their values $w_e$.  Note further
that since the $w_e$ do not affect the analysis, if our $w_e$ violate
the assumption that $w_e\le 1$, we can scale them all by dividing by
$\max w_e$ and apply the same result.  So the restriction $w_e \le 1$
was actually irrelevant.

The key feature of this scheme is that an item's greater weight is
translated into an increased probability of being sampled: this lets
it contribute more to the expectation of the sample without
contributing too much to its variance.

One might object that in order to apply the above scheme, we need to
know the weights $w_e$ in order to decide on the correct sampling
probabilities.  This would appear to imply a knowledge of the very
quantity we wish to compute.  It is at this point that we invoke the
specifics of our approach to avoid the difficulty.

We modify a uniform sampling scheme developed
previously~\cite{Karger:Skeleton}. That scheme sampled all graph edges
with the same probability and showed the following.

\begin{lemma}[\cite{Karger:Skeleton}]
\labelprop{Lemma}{prop:basic-sampling} Let $G$ be a graph in which ths
edges have mutually independent random weights, each distributed in
the interval $[0,1]$.  If the expected weight of every cut in $G$
exceeds $\rho_\epsilon=3(d+2)(\ln n)/\epsilon^2$ for some $\epsilon$
and $d$, then with probability $1-1/n^d$ every cut in $G'$ has value
within $(1\pm\epsilon)$ of its expectation.
\end{lemma}

The intuition behind this theorem is the same as for the Chernoff
bound.  In the sampled graph, the expected value of each cut is
$\Omega((\log n)/\epsilon^2)$, while each edge contributes value at
most 1 to the sampled cuts it is in.  Thus, the contribution of any
one edge to the possibile deviation of a cut from its mean is
negligible.\footnote{This theorem is nontrivial, as the exponential
number of cuts means that events which are very unlikely on one cut
still seem potentially probable over all cuts.  But it can be shown
that most cuts are so large in expectation that their deviation is
exponentially unlikely.}

As in our above discussion, we now observe that an edge that only
crosses large-valued cuts can have its sampled weight scaled up (and
its probability of being sampled correspondingly scaled down) without
making that edge dominate any of the samples it is in.  Consider a
$k$-connected induced subgraph of $G$ with
$k>c$. \ref{prop:basic-sampling} says that we can sample the edges of
this subgraph with probability $\O(1/k)$ (and scale their weights up by
$\O(k)$ to preserve expectations) without introducing significant
error in the cut values.  More generally, we can sample edges in any
subgraph with probability inversely proportional to the connectivity
of that subgraph.  We will generalize this observation to argue that
we can simultaneously sample each edge with probability inversely
proportional to the maximum connectivity of any subgraph containing
that edge.

To take advantage of this fact, we will show that almost all the edges
are in components with large connectivities and can therefore be
sampled with low probability---the more edges, the less likely they
are to be sampled.  We can therefore construct an $O(n\log n)$-edge
graph that, regardless of the minimum cut value, accurately
approximates all cut values.

\subsection{Definitions}
\label{sec:def}

We use the term ``unweighted graph'' to refer to a graph in which all
edges have weight 1.  In the bulk of this paper, $G$ denotes an
unweighted undirected graph with $n$ vertices and $m$ edges; parallel
edges are allowed. We also consider weighted graphs.  By scaling
weights, we can assume the minimum edge weight is at least one.  For
the purpose of correctness analysis when running times are not
relevant, it is often convenient to treat an edge of weight $w$ as a
set of $w$ parallel edges with the same endpoints.

A \emb{cut} $\C$ is a partition of the vertices into two subsets. The
\emb{value} $\val(\C,G)$ of the cut in unweighted (resp. weighted)
graph $G$ is the total number (resp. weight) of edges with endpoints
in different subsets.

We simplify our presentation with a vector notation.  The term $x_E$
denotes a vector assigning some value $x_e$ to each $e \in E$.
\emph{All} operations on vectors in this paper are
\emph{coordinatewise}.  The interpretation of $x_E+y_E$ is standard,
as is the product $\gamma x_E$ for any constant $\gamma$.  However, we
let $\gmul{x_E}{y_E}$ denote the product $z_E$ with $z_e=x_ey_e$.
Similarly, let $1/x_E$ denote the vector $z_E$ such that $z_e=1/x_e$
(pointwise inverse).  More generally, let $y_E/x_E$ be the vector
$z_E$ with $z_e=y_e/x_e$.

A weighted graph $G$ can be thought of as the vector (indexed by edge
set $E$) of its edge weights. (An unweighted graph has value 1 in all
coordinates.)  Applying our vector notation, when $r_E$ is a vector
over the edge set, we let $\gmul{r_E}{F}$ denote a graph with edge
weight vector $r_E F$.  Similarly, if $G$ and $H$ are graphs, then
$G+H$ denotes the graph whose edge weight vector is the sum of those
graphs'.

We also introduce a sampling notation.  As is traditional, we let
$\gsamp{p}{G}$ denote a graph in which each edge of $G$ is
incorporated with probability $p$.  Generalizing, we let
$\gsamp{p_E}{G}$ denote a random subgraph of $G$ generated by included
each edge $e$ of $G$ (with its original weight) independently with
probability $p_e$.  We define the \emph{expected value graph}
$E[\gsamp{p_E}{G}]=\gmul{p_E}{G}$, since the expected value of any
edge in $\gsamp{p_E}{G}$ is equal to the value of that edge in
$\gmul{p_E}{G}$.  This means that expected cut values are also
captured by the expected value graph.

We say that an event occurs with \emb{high probability} if its
probability is $1-O(n^{-d})$ for some constant $d$.  The constant can
generally be modified arbitrarily by changing certain other constants
hidden in the asymptotic notation.

\subsection{Outline}

In Section~\ref{sec:compression} we define the \emph{strong
connectivity} measure that is used to determine the relative impact of
different edges on cut samples, and show that samples based on this
strong connectivity measure have good concentration near their mean.
Our application to $s$-$t$ min-cuts is immediate.  In
Section~\ref{sec:smoothing} we introduce \emph{graph smoothing}, a
variation on compression that can be used for flow approximation.
Finally, in Section~\ref{sec:find-strength}, we show how the strong
connectivities needed for our sampling experiments can actually be
estimated quickly.

\section{Approximating Cuts via Compression}
\label{sec:compression}

As was stated above, we aim to sample edges with varying
probabilities.  To preserve cut values, we compensate for these
varying sampling probabilities using \emph{compression}.  To define
the appropriate sampling probability for each edge, we introduce the
notion of \emph{strong connectivity}.  For the bulk of this section,
we will focus on unweighted graphs, though we will occasionally make
reference to edge weights for future use.

\subsection{Compression}

Sampling edges with different probabilities means that cut values no
longer scale linearly. To make the expected cut value meaningful, we
counterbalance the varying sampling probabilities by introducing edge
weights on the sampled edges.

\begin{definition}
  Given an unweighted graph $G$ and {\em compression probabilities}
  $p_e$ for each edge $e$, we build a {\em compressed graph}
  $\gcomp{p_E}{G}$ by including edge $e$ in $\gcomp{p_E}{G}$ with probability
  $p_e$, and giving it weight $1/p_e$ if it is included.
\end{definition}

In our notation above, the compressed graph
$\gcomp{p_E}{G}=\gsamp{p_E}{\gmul{1/p_E}{G}}$.  Since the expected weight of
any edge in the graph is $1$, every cut's expected value is equal to
its original value, regardless of the $p_e$.  That is,
$E[\gsamp{p_E}{\gmul{1/p_E}{G}}]=G$.  However, the expected number of
edges in the graph is $\sum p_e$.  We would therefore like to make all
the $p_e$ as small as possible.  We are constrained from doing so,
however, by our need to have all the cut values tightly concentrated
around their expectations.  An edge compressed with probability $p_e$
has variance $(1-p_e)/p_e$, and the large variances produced by small
$p_e$ work against our wish for tight concentration.  The key
question, then, is how small we can make our $p_e$ values (and thus
our expected number of sampled edges) while preserving tight
concentration of cut values.

\subsection{Strong Connectivity}
\label{sec:strong-connectivity}

In this section, we formalize the notion of subgraphs with large
connectivities.  As was discussed above, if we identify a subgraph
with connectivity $k\gg c$, then we might hope, based on
\ref{prop:basic-sampling}, to sample edges in this subgraph with
probability roughly $1/k$, producing a graph much sparser than if we
sample with probability $1/c$.

\begin{definition}
  A graph $G$ is {\em $k$-connected} if the value of each cut in $G$
  is at least $k$.
\end{definition}
\begin{definition}
  A {\em $k$-strong component} of $G$ is a maximal $k$-connected
  vertex-induced subgraph of $G$.
\end{definition}

It follows that the $k$-strong components partition the vertices of a
graph and each $(k+1)$-strong component is contained in a single
$k$-strong component---that is, that the partition into $(k+1)$-strong
components refines the partition into $k$-strong components.

\begin{definition}
  The {\em strong connectivity} or \emph{strength} of an edge $e$,
  denoted $k_e$, is the maximum value of $k$ such that a $k$-strong
  component contains (both endpoints of) $e$.  We say $e$ is {\em
  $k$-strong} if its strong connectivity is $k$ or more, and {\em
  $k$-weak} otherwise.
\end{definition}

Note that the definition of strong connectivity of an edge differs
from the standard definition of connectivity:

\begin{definition}
  The {\em (standard) connectivity} of an edge $e$ is the minimum
  value of a cut separating its endpoints.
\end{definition}

Consider the graph with unit-weight edges $(s,v_i)$ and $(v_i,t)$ for
$i=1,\ldots,n$.  Vertices $s$ and $t$ have (standard) connectivity $n$
but only have strong connectivity $1$.  An edge's strong connectivity
is always less than its connectivity since an edge in a $k$-strong
component cannot be separated by any cut of value less than $k$.

\subsection{The Compression Theorem}

We now use the above definitions to describe our results.  We will use
a fixed \emb{compression factor} $\rho_\epsilon$ chosen to satisfy a
given error bound $\epsilon$:
\[
\const_\epsilon=3(d+4)(\ln n)/\epsilon^2\ .
\]

\begin{theorem}[Compression] Let $G$ be an unweighted graph with edge
  strengths $k_e$.  Given $\epsilon$ and a corresponding
  $\const_\epsilon$, for each edge $e$, let
  $p_e=\min\{1,\const/k_e\}$.  Then with probability $1-n^{-d}$,
\begin{enumerate}
\item The graph $\gcomp{p_E}{G}$ has $O(n\const)$ edges, and
\item every cut in $\gcomp{p_E}{G}$ has value between $(1-\epsilon)$ and
  $(1+\epsilon)$ times its value in $G$.
\end{enumerate}
\end{theorem}

In particular, to achieve any constant error in cut values with high
probability, one can choose $\const$ to yield $O(n\log n)$ edges in
the compressed graph.

We now embark on a proof of the Compression Theorem.  

\subsubsection{Bounding the number of edges}

To prove the first claim of the Compression Theorem we use the
following lemma:

\begin{lemma}
\labelprop{Lemma}{prop:convergent-sum}
In a weighted graph with edge weights $u_e$ and strengths $k_e$, 
\[
\sum u_e/k_e \le n-1.
\]
\end{lemma}
\begin{proof}
Define the \emph{cost} of edge $e$ to be $u_e/k_e$.  We show that the
total cost of edges is at most $n-1$.  Let $C$ be any connected
component of $G$ and suppose it has connectivity $k$.  Then there is a
cut of value $k$ in $C$.  On the other hand, every edge of $C$ is in a
$k$-strong subgraph of $G$ (namely $C$) and thus has strength at least
$k$.  Therefore,
\begin{eqnarray*}
\sum_{e \mbox{ crossing } C} u_e/k_e &\le &\sum u_e/k\\
&= &k/k\\
&= &1
\end{eqnarray*}
Thus, by removing the cut edges, of total cost at most 1, we can break
$C$ in two, increasing the number of connected components of $G$ by
1.  

If we find and remove such a cost-1 cut $n-1$ times, we will have a
graph with $n$ components.  This implies that all vertices are
isolated, meaning no edges remain.  So by removing $n-1$ cuts of cost
at most 1 each, we have removed all edges of $G$.  Thus the total cost
of edges in $G$ is at most $n-1$.
\end{proof}

This lemma implies the first claim of the Compression Theorem.  In our
graph compression experiment, all edge weights are one, and we sample
each $e$ with probability $\rho/k_e$.  It follows that the expected
number of edges is $\rho \sum 1/k_e \le \rho(n-1)$ by the previous
lemma.  The high probability claim follows by a standard Chernoff
bound~\cite{Chernoff,Motwani:RandomizedAlgorithms}.  

\subsubsection{Proving cuts are accurate} 

We now turn to the proof that cuts are accurate in the compressed
graph.  Once again, we apply a useful property of edge strengths.

\begin{lemma}
\labelprop{Lemma}{prop:unit-cut}
  If graph $G$ has edge strengths $k_e$ then the graph
  $\gmul{1/k_E}{G}$ has   minimum cut exactly 1.
\end{lemma}
\begin{proof}
  Consider any minimum cut in $G$, of value $c$.  Each edge in the cut
  has strength $c$, giving it weight $1/c$ in $\gmul{1/k_E}{G}$.
  Thus, the cut has value 1 in $\gmul{1/k_E}{G}$.  It follows that the
  minimum cut in $\gmul{1/k_E}{G}$ is at most $1$.
  
  Now consider any cut, of value $k$ in $G$.  Each edge crossing the
  cut has strength at most $k$, meaning it gets weight at least $1/k$
  in $\gmul{1/k_E}{G}$.  Since $k$ edges cross this cut, it follows
  that the cut has weight at least $k(1/k) \ge 1$.  This shows that
  the minimum cut in $\gmul{1/k_E}{G}$ is at least 1.

  Combining these two arguments yields the claimed result.
\end{proof}

Recall that for graph compression, we initially assign weight $k_e$ to
edge $e$, producing a weighted graph $\gmul{k_E}{G}$.  We then produce
a random graph by choosing edge $e$ of $\gmul{k_E}{G}$ with
probability $\rho/k_e$, generating the graph
$\gsamp{\rho/k_E}{\gmul{k_E}{G}}$ (we assume for the moment that all
$k_e \ge \rho$ so the sampling probability is at most 1).  Our goal is
to show that the resulting graph has cuts near their expected values.

Our basic approach is to express $\gmul{k_E}{G}$ as a weighted sum of
graphs, each of which, when sampled, is easily proven to have cut
values near their expectations.  It will follow that the sampled
$\gsamp{\rho/k_E}{\gmul{k_E}{G}}$ also has cut values near its
expectations.

We now define the decomposition of $G$.  There are at most $m$
distinct edge-strength values in $G$, one per edge (in fact it can be
shown there are only $n-1$ distinct values, but this will not matter).
Number these values $k_1,\ldots,k_r$ in increasing order, where $r \le
m$.  Now define the graph $F_i$ to be the edges of strength at least
$k_i$---in other words, $F_i$ is the set of edges in the $k_i$-strong
components of $G$.  Write $k_0=0$.  We now observe that
\[
\gmul{k_E}{G} = \sum_i \gmul{(k_{i}-k_{i-1})}{F_i}.
\]
To see this, consider some edge of strength exactly $k_i$.  This edge
appears in graphs $F_1, F_{2}, \ldots, F_i$.  The total weight
assigned to that edge in the right hand of the sum above is therefore
\[
  (k_1-k_0) + (k_2-k_1) + \cdots (k_i-k_{i-1}) = k_i-k_0 = k_i
\]
as is required to produce the graph $\gmul{k_E}{G}$ which has weight
$k_e$ on edge $e$.
  
We can now examine the effect of compressing $G$ by examining its
effect on the graphs $F_i$.  Our compression experiment flips an
appropriately biased coin for each edge of $\gmul{k_E}{G}$ and keeps
it if the coin shows heads.  We can think of these coin flips as also
being applied to the graphs $F_i$.  We apply the same coin flip to all
the $F_i$: edge $e$ of strength $k_i$, present in $F_1,\ldots,F_i$, is
kept in all of the respective samples $\gsamp{\rho/k_E}{F_i}$ if the
coin shows heads, it is discarded from all if the coin shows tails.
Thus, the samples from the graphs $F_i$ are not independent.  However,
if we consider a \emph{particular} $F_i$, then the sampling outcomes
of edges are mutually independent \emph{in that particular $F_i$}.

Let us first consider graph $F_1$ (which is simply the graph $G$ since
all $k_e \ge 1$).  As was discussed in Section~\ref{sec:def}, the
expected value $E[\gsamp{\rho/k_E}{G}]=\gmul{\rho/k_E}{G}$ has cut
values equal to the expectations of the corresponding cuts of the
sampled graph $\gsamp{\rho/k_E}{G}$.  We saw above that the graph
$\gmul{1/k_E}{G}$ has 
minimum cut 1.  It follows that the expected value graph
$\gmul{\rho/k_E}{G}$ has minimum cut $\rho$.  This suffices to let us
apply the basic sampling result (\ref{prop:basic-sampling}) and deduce
that every cut in $F_1$ has value within $(1\pm\epsilon)$ of it
expectation with high probability.  Scaling the graph preserves this:
the graph $\gsamp{1/k_E}{\gmul{(k_1-k_0)}{F_1}}$ has cut values within
$(1\pm\epsilon)$ of their expectations with high probability.
  
Now consider any other $F_i$.  The subgraph $F_i$ consists of all the
edges inside the $k_i$-strong components of $G$.  Consider one
particular such component $C$, and an edge $e \in C$.  Since $C$ is
$k_i$-connected, we know that $k_e \ge k_i$.  By definition, edge $e$
is contained in some $k_e$-connected subgraph of $G$.  As was argued
above in Section~\ref{sec:strong-connectivity}, the $k_e$-connected
subgraph that contains $e$ must be wholly contained in $C$.  Thus, the
strength of edge $e$ \emph{with respect to the graph $C$} is also
$k_e$.\footnote{%
This proof step is the sole motivation for the introduction
of strong connectivity.  The nesting of strong components lets us draw
conclusions about the graphs $F_i$ that \emph{cannot} be drawn about
standard connectivity.  The set of edges with standard connectivity
exceeding $k$ does \emph{not} form a $k$-connected graph, which
prevents our proof from going through when we use standard
connectivity.

Nonetheless, it is conceivable that standard connectivity is a
sufficient metric for our sampling algorithm.  We have found no
counterexample to this possibility.
}
Our argument of the previous paragraph for graph $G$ therefore applies
to the graph $C$, implying that the sampled version of $C$ in
$\gsamp{\rho/k_E}{F_i}$ has cuts within $(1\pm\epsilon)$ of their
expected values with high probability.  Since this is true for each
component $C$, it is also true for the graph $F_i$ (since each cut of
$F_i$ is a cut of components of $F_i$).
  
This completes our argument.  We have shown that each
$\gsamp{\rho/k_E}{F_i}$ has all cuts within $(1\pm\epsilon)$ of their
expected values with probability $1-1/n^{d+2}$ (the quantity $d+2$
follows from our choice of $\rho$ and the application of
\ref{prop:basic-sampling}).  Even though the $\gsamp{1/k_E}{F_i}$ are
not independent, it follows from the union bound that all (possibly
$n^2$) distinct $F_i$ samples are near their expectation with
probability $1-1/n^d$.  If this happens, then the sample
$\gmul{k_E}{\gsamp{1/k_E}{G}} = \sum
\gmul{(k_i-k_{i-1})}{\gsamp{1/k_E}{F_i}}$ has all cuts within
$1\pm\epsilon$ of their expected values (this follows because all
multipliers $k_i-k_{i-1}$ are positive).  Of course, the expected
graph $E[\gmul{k_E}{\gsamp{1/k_E}{G}}]=G$.

Our analysis has assumed all edges are sampled with probability
$\rho/k_e$, which is false for edges with $k_e < \rho$ (their sampling
probability is set to 1 in the Compression Theorem).  To complete the
analysis, consider the $\rho$-strong components of $G$.  Edges outside
these components are not sampled.  Edges inside the components are
sampled with probabilities at most 1.  We apply the argument above to
each $\rho$-strong component separately, and deduce that it holds for
the entire compressed graph.

\subsection{Weighted Graphs}

For simplicity, our compression analysis was done in terms of
unweighted graphs.  However, we can apply the same analysis to a
weighted graph.  If the weights are integers, we can think of a weight
$u$ edge as a set of $u$ parallel unit-weight edges and apply the
analysis above.  Given the strengths $k_e$, we would take each of the
$u$ edges with probability $1/k_e$ and give it weight $k_e$ if taken.
Of course, if $u$ is large it would take too much time to perform a
separate coin flip for each of the $u$ edges. However, we can see that
the number of edges actually taken has a binomial distribution with
parameters $u_e$ and $\rho/k_e$; we can sample directly from that
binomial distribution.  Note that the number of edges produced is
$O(n\log n)$ regardless of the $u_e$.

To handle noninteger edge weights, imagine that we multiply all the
edge weights by some large integer $z$.  This uniformly scales all the
cut values by $z$.  It also scales all edge strengths by $z$.  If we
now round each edge down to the nearest integer, we introduce an
additive error of at most $m$ to each cut value (and strength); in the
limit of large $z$, this is a negligible relative error.  To compress
the resulting graph, the approach of the previous paragraph now says
that for a particular edge $e$ with original weight $u_e$, we must
choose from a binomial distribution with parameters $\floor{zu_e}$
(for the number of edges, which has been multiplied by $z$ and
rounded) and $\rho/zk_e$ (since all edges strengths have also been
multiplied by $z$).  In the limit of large $z$, it is well
known~\cite{Feller} that this binomial distribution converges to a
\emph{Poisson Distribution} with parameter $\lambda = \rho u_e/k_e$.
That is, we produce $s$ sample edges with probability
$e^{-\lambda}\lambda^s/s!$.   Under the compression formula,
their weights would each be $zk_e/\rho$.  Recall, however that we
initially scaled the graph up by $z$; thus, we need to scale back down
by $z$ to recover $G$; this produces edge weights of $k_e/\rho$.

From an algorithmic performance perspective, we really only care
whether the number of sampled edges is 0 or nonzero since, after
sampling, all the sampled edges can be aggregated into a single edge
by adding their weights.  Under the Poisson distribution, the
probability that the number of sampled edges exceeds 0 is $1-e^{-\rho
u_e/k_e}\approx \rho u_e/k_e $.  It is tempting to apply this
simplified compression rule to the graph (take edge $e$ with
probability $\rho u_e/k_e$, giving it weight $k_e/\rho$ if taken).  A
generalized Compression theorem in the appendix shows that this
approach will indeed work.

\subsection{Using Approximate Strengths}

Our analysis above assumed edge strengths were known.  While edge
strengths can be computed exactly, the time needed to do so would make
them useless for cut and flow approximation algorithms.  Examining the
proofs above, however, shows that we do not need to work with exact
edge strengths.

\begin{definition}
Given a graph $G$ with $n$ vertices, edge weights $u_e$, and edge
strengths $k_e$, a set of edge value ${\tilde k}_e$ are \emph{tight
strength bounds} if
\begin{enumerate}
\item ${\tilde k}_e \le k_e$ and
\item $\sum u_e/{\tilde k}_e = O(n)$
\end{enumerate}
\end{definition}

\begin{theorem}
\labelprop{Theorem}{prop:tight-enough}
The Compression Theorem remains true even if tight strength bounds are
used in place of exact strength values.
\end{theorem}
\begin{proof}
The proof of cut accuracy relied on the fact that each sampled edge
had small weight compared to its cuts.  The fact that ${\tilde k}_e
\le k_e$ means that the weights of included edges are smaller than
they would be if true strengths were used, which can only help.

The bound on the number of edges in the compressed graph followed
directly from the fact that $\sum u_e/k_e \le n$; for tight strength
bounds this summation remains asymptotically correct.
\end{proof}

Tight strength bounds are much easier to compute than exact strengths.

\begin{theorem}
\labelprop{Theorem}{prop:find-strengths}
Given any $m$-edge, $n$-vertex graph, tight strength bounds can be
computed in $O(m\log^2 n)$ time for unweighted graphs and $O(m\log^3
n)$ time for weighted graphs.
\end{theorem}
\begin{proof}
See Section~\ref{sec:find-strength}.
\end{proof}

\subsection{Applications}

We have shown that graphs can be compressed based on edge strengths
while preserving cut values.  This suggests that cut problems can be
approximately solved by working with the compressed graph as a
surrogate for the original graph.  We now prove the application
corollaries from the introduction.

\subsubsection{{\boldmath Minimum $s$--$t$ cuts.\quad}}

As discussed above, we can compute tight strengths bounds in
$\Olog(m)$ time and generate the resulting compressed graph
$\gcomp{p_E}{G}$ as described in the Compression Theorem.  The graph
will have $O(\rho n)=O(n(\log n)/\epsilon^2)$ edges.

Let us fix a pair of vertices $s$ and $t$. Let $\hat v$ be the value
of a minimum cut separating $s$ from $t$ in the compressed graph
$\gcomp{p_E}{G}$. We show that the minimum $s$--$t$ cut value $v$ in
$G$ is within $(1\pm 3\epsilon)\hat v$. By the Compression Theorem,
with high probability the $s$--$t$ minimum cut $\C$ in $G$ has value
at most $(1+\epsilon) v$ in $\gcomp{p_E}{G}$. Thus $\hat v \le
(1+\epsilon)v$.  Furthermore, with high probability every cut of $G$
with value exceeding $(1+3\epsilon)v$ in $G$ will have value at least
$(1-\epsilon)(1+3\epsilon) \ge (1+\epsilon)v$ in $\gcomp{p_E}{G}$ and
therefore will not be the minimum cut of $\gcomp{p_E}{G}$.

We can find an approximate value $\hat v$ of the minimum $s$--$t$ cut
(and an $s$--$t$ cut with this value) by computing a maximum flow in
the $O(n\log n\,/\epsilon^2)$-edge graph $\gcomp{p_E}{G}$. The maximum
flow algorithm of Goldberg and Tarjan~\cite{Goldberg:Maxflow} has a
running time of $O(nm\log(n^2/m))$ which leads to a running time
of $O(n^2\log^2n\,/\epsilon^2)$ after compression.  Similarly, the
Goldberg-Rao algorithm~\cite{Goldberg:CapacitatedBlocking}, which runs
in $\Olog(m^{3/2})$ time, leads to a running time of
$\Olog(n^{3/2}/\epsilon^3)$ after compression.  

In an integer-weighted graph with small flow value, we may wish to
apply the classical augmenting path
algorithm~\cite{Ford:Maxflow,Ahuja:Flow} that finds a flow of value
$v$ in $v$ augmentations.  As described, the graph-compression process
can produce noninteger edge weights $\rho/k_e$, precluding the use of
augmenting paths in the smoothed graph.  However, if we decrease each
compression weight to the next lower integer (and increase the
sampling probability by an infinitesimal amount to compensate) then
compression will produce an integer-weighted graph in which the
augmenting paths algorithm can be applied to find an $s$--$t$ cut of
value at most $(1+\epsilon)v$ in time $O(nv\log n\,/\epsilon^2)$.

\subsubsection{Sparsest cuts}

A {\em sparsest cut} of a graph $G$ minimizes the ratio between the
cut value and the product of number of vertices on the two sides.  It
is $\NP$-hard to find the value of a sparsest cut. To find an
$\alpha$-approximate value of a sparsest cut, we use the approach of
the previous subsection: we compute a $\beta$-approximate sparsest cut
in the compressed graph $\gcomp{p_E}{G}$.  This cut is then an
$\alpha=(1+\epsilon)\beta$-approximate sparsest cut of $G$.

An algorithm of Klein, Stein and Tardos~\cite{Klein:Concurrent} finds
an $O(\log n)$-approximation to a sparsest cut in $O(m^2\log m)$
time. By running their algorithm on $\gcomp{p_E}{G}$, we will find an
$O(\log n)$-approximate sparsest cut in $O(n^2\log^3n\,/\epsilon^4)$
time.  Our small cut-sampling error is lost asymptotically in the
larger error of the 
approximation algorithm.

Our approach been applied in a similar way to improve the running
time of a spectral partitioning algorithm~\cite{Kannan:Clustering}.

\section{Approximating Flows by Graph Smoothing}
\label{sec:smoothing}

Until now we have focused on cut problems.  Our compression scheme
produces a graph with nearly the same cut values as the original, so
that cut problems can be approximated in the compressed graph.  But
consider a maximum flow problem.  It would seem natural to try to
approximate this maximum flow by finding a maximum flow in the
compressed graph.  By providing an approximately minimum $s$-$t$ cut,
this approach does indeed give an approximation to the \emph{value} of
the maximum flow.  But since edges in the compressed graph have
\emph{larger} capacity than the original graph edges, a feasible flow
in the compressed graph will probably not be feasible for the original
graph.

Previous work~\cite{Karger:Skeleton} tackled the flow approximation
problem with a divide-and-conquer approach.  The edges of $G$ are
randomly divided into a number of groups, producing several random
subgraphs of $G$.  \ref{prop:basic-sampling} is applied to deduce that
each subgraph has cut values near their expectations.  By computing a
flow in each subgraph and adding the flows, we find a flow of value
$(1-\epsilon)$ times the maximum flow in $G$.  

This approach suffers the same limitation as the uniform sampling
approach for cuts: the probability of each edge occurring in each
subgraph must be $\Omega(1/c)$ to preserve cut values.  This
translates into a limit that we divide into $O(c)$ groups, which
limits the power of the scheme on a graph with small minimum cuts.
Graph compression's nonuniform sampling approach does not seem to
provide an immediate answer: clearly we cannot simultaneously divide
each edge with strength $k_e$ among $k_e$ distinct subgraphs.  Instead
we need a consistent rule that divides all edges among a fixed number
of subgraphs.  Each subgraph must therefore look like a \emph{uniform}
sample from the original graph.

In this section we introduce {\em graph smoothing}---a technique that
lets us apply uniform sampling, and through it analyze randomized
divide and conquer algorithms, for graphs with small minimum cuts,
yielding fast approximation algorithms for flows in such graphs.  The
approach applies equally well to weighted graphs.

Our approach again starts with \ref{prop:basic-sampling}. The sampling
proof used a Chernoff bound, which relied on individual edges having
only a small impact on the outcome of the experiment.  In particular,
since the graph had minimum cut $c$, and every edge was being chosen
with probability $p$, every cut had expected value at least $pc$.
Thus, the presence or absence of a single (weight 1) edge could
affect that value of a cut by at most a $1/pc$-fraction of its
expected value.    

If we want to be able to sample more sparsely, we run into a problem
of certain edges contributing a substantial fraction of the expected
value of the cuts they cross, so that the Chernoff bound breaks down.
A fix is to divide such edges into a number of smaller-weight
edges so that they no longer dominate their cuts.  Dividing \emph{all}
the graph edges is quite pointless: splitting all edges in half has
the effect of doubling the minimum cut (allowing us to sample at half
the original rate while preserving approximate cut values), but since
we double the number of edges, we end up with the same number of
sampled edges as before.

The approach of $k$-strong components lets us circumvent this problem.
We use $k$-strong components to show that only a small fraction of the
graph's edges are large compared to their cuts.  By dividing
\emph{only} those edges, smoothing the highest-variability features of
the sample, we allow for a sparser sample that still preserves cut
values.  Since only a few edges are being divided, the random
subgraphs end up with fewer edges than before, making algorithms based
on the samples more efficient.

\subsection{Smooth Graphs}

For the study of graph compression, we focused on unweighted graphs.
For smoothing we focus on weighted graphs.  In keeping with standard
terminology for flows, we will refer to weights as \emph{capacities}.
It is easy to extend the notation $\gsamp{p}{G}$ to denote taking each
capacitated edge with probability $p$, but somewhat harder to prove
that sampling does the right thing.  As discussed above, the problem
is that a single capacitated edge might account for much of the
capacity crossing a cut.  The presence or absence of this edge has a
major impact on the value of this cut in the sampled graph.  However,
the idea of edge strength described above gives us a useful bound on
how much impact a given edge can have.

\begin{definition}
A graph $G$ with edge capacities $u_e$ and edge strengths $k_e$ is
{\em $c$-smooth} if for every edge, $k_e \ge c u_e$.
\end{definition}

Note that a graph with integer edge weights and minimum cut $c$ has
smoothness at most $c$ but possibly much less.  We now argue that
smoothness is the criterion we need to apply uniform sampling to
weighted graphs.

\begin{theorem} 
\labelprop{Theorem}{prop:smooth sample} Let $G$ be a $c$-smooth graph.
Let $p=\rho_\epsilon/c$ where $\rho_\epsilon=O((\log n)/\epsilon^2)$
as in the Compression Theorem.  Then with high probability, every cut
in $\gsamp{p}{G}$ has value in the range $(1\pm\epsilon)$ times its
expectation (which is $p$ times its original value).
\end{theorem}
\begin{proof}
We use a variation on the proof of the Compression Theorem.  Given the
graph $G$, with edge capacities $u_e$, let $k_i$ be a list of the at
most $m$ strengths of edges in $G$ in increasing order, and let $F_i$
denote the graph whose edge set is the $k_i$-strong edges of $G$, but
with edge $e$ assigned weight $c u_e/k_e$.  It follows, just as was
argued above, that $G=\sum (k_i-k_{i-1}) F_i$.  So if we prove that
each $F_i$ can be accurately sampled with probability $p=\rho/c$, then
the same will apply to $G$.

So consider graph $F_i$.  Since we have assigned weights $c u_e/k_e$,
the minimum cut in $F_i$ is $c$, as was argued in \ref{prop:unit-cut}.
At the same time, edge $e$, if present in this graph, has weight
$cu_e/k_e \le 1$ by the smoothness property.  It follows that we can
apply \ref{prop:basic-sampling} to each component of the graph $F_i$
and deduce that all cuts are within $(1\pm\epsilon)$ of their
expectation, as desired.  The remainder of the proof goes as for the
Compression Theorem.
\end{proof}

\subsection{Making Graphs Smooth.}
\label{sec:make smooth}

We have shown that a smooth graphs can be sampled uniformly, which
will lead to good flow algorithms.  We now give algorithms for
transforming any graph into a smooth one.

\begin{lemma}
\labelprop{Lemma}{prop:make-smooth}
  Given an $m$ edge capacitated graph, a {\em smoothness parameter}
    $c$ and the strengths $k_e$ of all edges, we can transform the
    graph into an $m+cn$-edge $c$-smooth graph in $\Olog(m)$ time.
\end{lemma}
\begin{proof}
Divide edge $e$ into $\ceil{c u_e/k_e}$ parallel edges, each of
capacity $u_e/\ceil{c u_e/k_e} \le k_e/r$ but with total capacity
$u_e$.  These edges remain $k_e$ strong, but now satisfy the
smoothness criterion.

It remains to prove that this division creates at most $nr$ new edges.
The number of edges in our smoothed graph is
\begin{eqnarray*}
\sum_e \ceil{c u_e/k_e}  &\le &m+\sum c u_e/k_e\\
&= &m+c \sum u_e/k_e\\
&\le &m+cn
\end{eqnarray*}
where the last line follows from \ref{prop:convergent-sum}.
\end{proof}

\begin{corollary}
Given edge strengths, in $O(m)$ time we can transform any $m$-edge
capacitated graph into an $O(m)$-edge capacitated $(m/n)$-smooth
graph.
\end{corollary}

Choosing the smoothness parameter $m/n$ is in some sense optimal.  Any
smaller smoothness parameter leads to worse sampling performance
without decreasing the asymptotic number of edges (which is always at
least $m$).  A larger smoothness parameter provides better sampling
behavior, but linearly increases the number of edges such that the
gains from sparser sampling are lost.

\subsection{Approximate Max-Flows}

To approximate flows, we use the graph smoothing technique.  As was
argued in \ref{prop:tight-enough}, graph smoothing works unchanged
even if we use tight strength bounds, rather than exact strengths, in
the computation.

After computing tight strength bounds in $\Olog(m)$ time (as will be
discussed in Section~\ref{sec:find-strength}), we can apply
\ref{prop:smooth-approx}. This shows that in any $c$-smooth graph,
sampling with probability $p$ produces a graph in which with high
probability all cuts are within $(1 \pm\epsilon)$ of their expected
values.  This fact is the only one used in the uncapacitated graph
flow algorithms of~\cite{Karger:Skeleton}.  Therefore, those results
immediately generalize to the smooth graphs defined here---we simply
replace ``minimum cut'' with ``smoothness'' in all of those results.
The generalization is as follows:

\begin{lemma}
\labelprop{Lemma}{prop:smooth-approx} Let $T(m,n,v,c)$ be the time to
find a maximum flow in a graph with $m$ edges, $n$ vertices, flow $v$
and smoothness $c$.  Then for any $\epsilon$, the time to find a flow
of value $(1-\epsilon)v$ on an $m$-edge, $n$-vertex, smoothness-$c$
graph is
\[
\Olog(\frac{1}{p}T(pm,n,pv,pc))
\]
where $p=\Theta((\log n)/\epsilon^2 c)$.
\end{lemma}
\begin{proof} Divide the graph edges into $1/p$ random
groups.  Each defines a graph with $pm$ edges.  Since the minimum
$s$-$t$ cut of $G$ is $v$, the minimum expected $s$-$t$ cut in each
group is $pv$.  By the Smoothing Theorem, each sample has minimum
$s$-$t$ cut, and thus maximum $s$-$t$ flow, at least $(1-\epsilon)pv$.
Find flows in each piece, and combine the results.  This total flow
will be $(1/p)(1-\epsilon)pv=(1-\epsilon)v$.
\end{proof}

\begin{corollary}
In any undirected graph, given edge strengths, a $(1-\epsilon)$-times
maximum flow can found in $\Olog(m\sqrt{n}/\epsilon)$ time.
\end{corollary}
\begin{proof}
Begin by converting the graph to an $O(m)$-edge $(m/n)$-smooth graph,
as discussed in \ref{prop:make-smooth}.  The Goldberg-Rao flow
algorithm~\cite{Goldberg:CapacitatedBlocking} gives
$T(m,n)=\Olog(m^{3/2})$ for the previous lemma.  (Since we are already
giving up a factor of $\epsilon$, we can assume without loss of
generality that all edge capacities are polynomial, thus eliminating
the capacity scaling term in their algorithm.)  Plugging this in gives
a time bound of $\O(m\sqrt{n}/\epsilon)$.
\end{proof}

Unlike for minimum cuts, it is not possible to use the standard
augmenting paths algorithm to find a flow in $\Olog(nv/\epsilon^2)$
time.  The graph smoothing process would subdivide unit-cost edges,
producing variable cost edges to which unit-capacity augmenting flows
cannot be applied.

In previous work~\cite{Karger:SmoothFlow}, Karger used the above techniques
to compute exact flows more quickly than before; however, this work
has been superseded by better algorithms (also based on edge
strength)~\cite{Karger:ResidualFlow}.

\section{Finding strong connectivities}
\label{sec:find-strength}

To efficiently compress and smooth graphs we would like to efficiently
find the strong connectivities of edges.  Unfortunately, it is not
clear that this can be done ($n$ maximum flow computations are one
slow solution).  But as discussed in \ref{prop:tight-enough}, we do
not require the exact values $k_e$.  We now show that it is possible
to find tight strength bounds ${\tilde k}_e$ that satisfy the two key
requirements of that Theorem: that ${\tilde k}_e \le k_e$ and $\sum
1/{\tilde k}_e =O(n)$.  These suffice for the cut and flow algorithms
described above.

Our basic plan begins with the following lemma.

\begin{lemma}
\labelprop{Lemma}{prop:kstrong} The total weight of a graph's $k$-weak
edges is at most $k(n-1)$.  In particular, any unweighted graph with
more than $k(n-1)$ edges has a nontrivial $k$-strong component (which
may be the entire graph).
\end{lemma}
\begin{proof}
Let $S$ be the set of $k$ weak edges, and suppose that the total
weight of edges in $S$ exceeds $k(n-1)$.  Then
\begin{align*}
\sum u_e/k_e & \ge \sum_{e \in S} u_e/k_e\\
& > \sum_{e \in S} u_e/k\\
& > k(n-1)/k\\
&= n-1
\end{align*}
which contradicts \ref{prop:convergent-sum}.
\end{proof}

We apply this lemma first to unweighted graphs.  \ref{prop:kstrong}
says that any unweighted graph with $k(n-1)$ or more edges has a
$k$-strong component.  It follows that at most $k(n-1)$ edges are
$k$-weak (that is, have strong connectivity less than $k$).  For
otherwise the subgraph consisting of the $k$-weak edges would have a
$k$-strong component, a contradiction.  For each value
$k=1,2,4,8,\ldots,m$, we will find a set of $k(n-1)$ edges containing
all the $k$-weak edges (note that every edge is $m$-weak).  We set
$k_e=k/2$ for all edges that are in the $k$-weak set but not the
$k/2$-weak set, thus establishing lower bounds for which the
Compression Theorem works.  The expected number of edges sampled under
this basic scheme would be
\[
\sum_{i=0}^{\log m} 2^i(n-1) (\const/2^i) = O(\const n \log m).
\]
We will eventually describe a more sophisticated scheme that
eliminates the factor of $\log m$.  It will also let us handle
weighted graphs efficiently.

\subsection{Sparse Certificates}
\label{sec:certificates}

A basic tool we use is {\em sparse certificates} defined by Nagamochi
and Ibaraki~\cite{Nagamochi:Connectivity}.

\begin{definition}
  A {\em sparse $k$-connectivity certificate,} or simply a {\em
    $k$-certificate}, for an $n$-vertex graph $G$ is a subgraph $H$ of
  $G$ such that
\begin{enumerate}
\item $H$ has $k(n-1)$ edges, and
\item $H$ contains all edges crossing cuts of value $k$ or less.
\end{enumerate}
\end{definition}

The certificate edges are related to $k$-weak edges, but are not quite
equivalent.  Any edge crossing a cut of value less than $k$ is
$k$-weak, but certain $k$-weak edges will not cross any cut of value
less than $k$.  We will show, however, that by finding $k$-certificate
edges one can identify $k$-weak edges.

Nagamochi and Ibaraki gave an algorithm~\cite{Nagamochi:Connectivity}
that constructs a sparse $k$-connectivity certificate in $O(m)$ time
on unweighted graphs, independent of $k$. 

\subsection{Finding $k$-weak edges}

Although a sparse $k$-certificate contains all edges with {\em
standard} connectivity less than $k$, it need not contain all edges
with {\em strong} connectivity less than $k$, since some such edges
might not cross any cut of value less than $k$.  We must therefore
perform some extra work.  In Figure~\ref{fig:weakedges} we give an
algorithm {\tt WeakEdges} for identifying edges with $k_e<k$.  It uses
the Nagamochi-Ibaraki {\tt Certificate} algorithm as a subroutine.

\begin{figure}[htbp]
\boxtext{

\noindent {\tt {\bf procedure} WeakEdges($G,k$)}\par
\medskip

\quad{\bf do} $\log_2 n$ {\bf times}\par
\quad\quad$E'\leftarrow{}${\tt Certificate($G,2k$)}\par
\quad\quad {\bf output} $E'$\par
\quad\quad$G\leftarrow G-E'$\par
\quad{\bf end do}\par
}
\caption{Procedure {\tt WeakEdges} for identifying $k_e < k$}
\label{fig:weakedges}
\end{figure}

\begin{theorem}
\labelprop{Theorem}{prop:weakedge}
{\tt WeakEdges} outputs a set containing all the $k$-weak edges of $G$.
\end{theorem}

\begin{proof}
  First suppose that $G$ has no nontrivial $k$-strong components,
  i.e. that $k_e<k$ for all edges. Then by \ref{prop:kstrong}, there
  are at most $k(n-1)$ edges in $G$; hence at least half of the
  vertices have at most $2k$ incident edges (which define a cut of
  value at most $2k$ with a single vertex on one side).  In an
  iteration of the loop in {\tt WeakEdges}, these vertices become
  isolated after removing the sparse certificate edges.  We have thus
  shown that in a single loop iteration half of the non-isolated
  vertices of $G$ become isolated.  The remaining graph still has no
  $k$-strong edges, so we can repeat the argument.  Hence in $\log_2
  n$ rounds we isolate \sb{all} vertices of $G$, which can only be
  done by removing all the edges. Thus all the edges of $G$ are output
  by {\tt WeakEdges}.

  In the general case, let us obtain a new graph $H$ by contracting
  each $k$-strong component of $G$ to a vertex. Any sparse
  $2k$-certificate of $G$ contains the edges of a sparse
  $2k$-certificate of $H$ as well.  Thus by the previous paragraph,
  all edges of $H$ are output by {\tt WeakEdges}.  But these are all
  the $k$-weak edges of $G$.
\end{proof}

\subsection{Sparse partitions}
 
Algorithm {\tt WeakEdges} can clearly be implemented via $O(\log n)$
calls to the Nagamochi-Ibaraki {\tt Certificate} algorithm.  It
follows that it runs in $O(m\log n)$ time on unweighted graphs and
outputs a set of at most $k(n-1)\log n$ edges.\footnote{It also
follows that a $k\log n$ sparse-certificate will contain all $k$-weak
edges, so they can be found with a single {\tt Certificate}
invocation.  This gives a better running time.  Indeed, since the
Nagamich Ibaraki algorithm ``labels'' each edge with the value $k$ for
which it vanishes, we can use those labels (divide by $\log n$) as
strength lower-bounds, producing a complete result in $O(m+n\log n)$
time.  However, this approach produces an extra $\log n$ factor in the
edge bound (or worse in weighted graphs) that we have been unable to
remove.}  In this section, we eliminate a $\log n$ factor in this
approach by finding edge sets that are ``sparser'' than the
Nagamochi--Ibaraki certificate.

The first observation we use is that a given $k$-certificate $E'$ may
contain edges that are inside a connected component of $G-E'$. The
edges in $G-E'$ do not cross any cut of value at most $k$ (by
definition of a sparse certificate), so the same holds for any edge of
$E'$ whose endpoints are connected by a path in $G-E'$. We can
therefore remove any such edge from $E'$ and put it back in $G$
without affecting the correctness of the proof of \ref{prop:weakedge}.

We can find the specified reduced edge set by contracting all edges
not in $E'$, yielding a new graph $G'$.  This effectively contracts
all (and only) edges connected by a path in $G-E'$.  But now observe
that any edge crossing a cut of value at most $k$ in $G$ also crosses
such a cut in $G'$ since we contract no edge that crosses such a small
cut. Thus we can find all edges crossing a small cut via a certificate
in $G'$.  Since $G'$ has fewer vertices, the certificate has fewer
edges.  We can iterate this procedure until all edges in the
certificate cross some cut of value at most $k$ or until $G'$ becomes
a single vertex. In the latter case, the original graph is
$k$-connected, while in the former, if the current contracted graph
has $n'$ vertices, it has at most $k(n'-1)$ edges.  This motivates the
following definition:

\begin{defn}
  A {\em sparse $k$-partition}, or {\em $k$-partition}, of $G$ is a
  set $E'$ of edges of $G$ such that
  \begin{enumerate}
  \item $E'$ contains all edges crossing cuts of value $k$ or less in
    $G$, and
  \item If $G-E'$ has $r$ connected components, then $E'$ contains at
    most $2k(r-1)$ edges.
  \end{enumerate}
\end{defn}

In fact, the construction just described yields a graph with at most
$k(r-1)$ edges, but we have relaxed the definition to $2k(r-1)$
edges to allow for an efficient construction.

Procedure {\tt Partition} in Figure~\ref{fig:partition} outputs a
sparse partition. It uses the Nagamochi--Ibaraki {\tt Certificate}
algorithm and obtains a new graph $G'$ by contracting those edges not
in the certificate. It repeats this process until the graph is
sufficiently sparse.

\begin{figure}[htbp]
\boxtext{

\begin{tabbing}

else \= else \= else \= \kill

{\tt {\bf procedure} Partition($G,k$)}\\
\\
{\bf input:} An $n$-vertex $m$-edge graph $G$\\
\\
\>{\bf if} $m \le 2k(n-1)$ {\bf then}\\
\>\> {\bf output} the edges of $G$\\
\>{\bf else}\\
\>\> $E'\leftarrow{}${\tt Certificate($G,k$)}\\
\>\> $G'\leftarrow{}$contract all edges of $G-E'$\\
\>\> {\tt Partition}($G',k$)
\end{tabbing}
}
\caption{ {\tt Partition} finds low-connectivity edges}
\label{fig:partition}
\end{figure}

\begin{lemma}
\label{parsim}
{\tt Partition} outputs a sparse $k$-partition partition in $O(m)$
time on unweighted graphs.
\end{lemma}

\begin{proof} 
  Correctness is clear since no edge crossing a cut of value less than
  $k$ is ever contracted and at termination $m\le 2k(n-1)$; we need
  only bound the running time.  If initially $m<k(n-1)$ then the
  algorithm immediately terminates.  So we can assume $m \ge k(n-1)$.

  Suppose that in some iteration $m > 2k(n-1)$.  We find a sparse
  connectivity certificate with $m' \le k(n-1)$ edges and then
  contract the graph to $n'$ vertices.  If $n'-1>(n-1)/2$ then in the
  following iteration we will have $m' \le k(n-1) < 2k(n'-1)$ and the
  algorithm will terminate.  It follows that the number of vertices
  (minus one) halves in every recursive call except the last.

  A single iteration involves the $O(m)$-time sparse-certificate
  algorithm~\cite{Nagamochi:Connectivity}.  At each recursive call,
  the edges remaining are all $k$-certificate edges from the previous
  iteration.  The number of such certificate edges is at most $k$
  times the number of vertices---thus the (upper bound on the) number
  of edges halves in each recursive call.  It follows that after the
  first call we have $T(n) = O(kn)+T(n/2) = O(kn)$.  This is $O(m)$
  since $m \ge k(n-1)$ by assumption.
\end{proof}

\begin{lemma}
  If {\tt Partition} is used instead of {\tt Certificate} in a call to
  {\tt WeakEdges$(G,k)$} (meaning we invoke {\tt Partition}$(G,2k)$
  instead of {\tt Certificate}$(G,2k)$), then algorithm {\tt
  WeakEdges} runs in $O(m\log n)$ time on unweighted graphs and
  returns a partition of $G$ into $r$ components for some $r$.  There
  are at most $4k(r-1)$ cross-partition edges and they include all the
  $k$-weak edges of $G$.
\end{lemma}
Note that the partition output by {\tt WeakEdges} is itself almost a
sparse $k$-partition; it simply has twice as many edges as the
definition allows.  On the other hand, it contains all $k$-weak edges;
not just the ones crossing small cuts.
\begin{proof}
  The running time follows from the previous lemma.  To prove the edge
  bound, consider a particular connected component $H$ remaining in a
  particular iteration of {\tt WeakEdges}.  A call to {\tt
  Partition}$(H,2k)$ returns a set of $4k(s-1)$ edges that breaks that
  component into $s$ subcomponents (the multiplier 4 arises from the
  fact that we look for a $2k$-partition).  That is, it uses at most
  $4k(s-1)$ edges to increase the number of connected components by
  $s-1$.  We can therefore charge $4k$ edges to each of the new
  components that gets created.  Accumulating these charges over all
  the calls to {\tt Partition} shows that if {\tt WeakEdges} outputs
  $4k(r-1)$ edges then those edges must split the entire graph into at
  least $r$ components.
\end{proof}

\subsection{Assigning Estimates}

We now give an algorithm {\tt Estimation} in
Figure~\ref{fig:estimation} for estimating strong connectivities.  We
use subroutine {\tt WeakEdges} to find a small edge set containing all
edges $e$ with $k_e<k$ but replace the Nagamochi-Ibaraki {\tt
Certificate} implementation with our algorithm {\tt Partition} to
reduce the number of output edges.

We assign values ${\tilde k}_e$ as follows.  In the first step, we run
{\tt WeakEdges} on $G$ with $k=2$; we set ${\tilde k}_e=1$ for the
edges in the output edge set $E_0$.  Then we delete $E_0$ from $G$;
this breaks $G$ into connected components $G_1,\ldots,G_\ell$.  Note
that each edge in $G_i$ has $k_e\ge2$ in $G$, though possibly not in
$G_i$. Then we recursively repeat this procedure in each $G_i$, by
setting $k=4$ in {\tt WeakEdges} and labeling all output edges with
${\tilde k}_e=2$, then with $k=8,16,\ldots,m$.  At the $i^{th}$ step,
all as-yet unlabeled edges have $k_e \ge 2^i$; we separate all those
with $k_e < 2^{i+1}$ and give them (valid lower bound) label ${\tilde
k}_e=2^i$.  Thus we find all ${\tilde k}_e$-values in at most $\log m$
iterations since $m$ is the maximum strength of an edge in an
unweighted graph.

\begin{figure}[htbp]
\boxtext{
\begin{tabbing}
foo \= foo \= foo \= foo \= foo \= \kill 
 {\bf procedure}  {\tt Estimation}($H,k$)\\
\\
{\bf input:} subgraph $H$ of $G$\\
\\
$E' \leftarrow{}$ {\tt WeakEdges($H$,$2k$)}\\
{\bf for each} $e \in E'$\\
\> $k_e \leftarrow k$\\
{\bf for each} nontrivial connected component $H' \subset H-E'$\\
\> {\tt Estimation($H'$,$2k$)}
\end{tabbing}
}
\caption{Procedure {\tt Estimation} for assigning ${\tilde k}_e$-values}
\label{fig:estimation}
\end{figure}

\begin{lemma}
\labelprop{Lemma}{prop:estimation-correct} If $H$ is any subgraph of
$G$, then {\tt Estimation}$(H,k)$ assigns lower bounds ${\tilde k}_e
\le k_e$ for all edges $e\in H$ with $k_e \ge k$ in $G$.
\end{lemma}
\begin{corollary}
After a call to {\tt Estimation$(G,1)$}, all the labels ${\tilde k}_e$
satisfy ${\tilde k}_e \le k_e$.
\end{corollary}
\begin{proof}
We prove the lemma by induction on the size of $H$.  The base case of
a graph with no edges is clear.  To prove the inductive step we need
only consider edges $e$ with $k_e \ge k$.  We consider two
possibilities.  If $e$ is in the set $E'$ returned by {\tt
WeakEdges}$(H,2k)$ then it receives label $k$, which is a valid lower
bound for any edge with $k_e \ge k$.  So the inductive step is proved
for $e \in E'$.  On the other hand, if $e \notin E'$, then $e$ is in
some $H'$ upon which the algorithm is invoked recursively.  By the
correctness of {\tt WeakEdges} we know $k_e \ge 2k$ (in $H$, and thus
in $G$) in this case. Thus, the inductive hypothesis applies to show
that $e$ receives a valid lower bound upon invocation of {\tt
WeakEdges}$(H',2k)$.
\end{proof}

\begin{lemma}
  Assume that in procedure {\tt WeakEdges}, procedure {\tt
    Certificate} is replaced by {\tt Partition}. Then the values
    ${\tilde k}_e$ output by {\tt Estimation$(G,1)$} are such that $\sum
    1/{\tilde k}_e=O(n)$.
\end{lemma}
\begin{proof} 
The proof is similar to the proof that $\sum u_e/k_e\le n$.  Define
the cost of edge $e$ to be $1/{\tilde k}_e$.  We prove that the total
cost assigned to edges is $O(n)$.  Consider a call to {\tt
Estimation}$(H,k)$ on some remaining connected component of $G$.  It
invokes {\tt WeakEdges$(H,k)$}, which returns a set of $4k(r-1)$ edges
whose removal partitions $H$ into $r$ connected components.  (Note
that possibly $r=0$ if $H$ is $k$-connected.)  The algorithm assigns
values ${\tilde k}_e=k$ to the removed edges.  It follows that the
total \emph{cost} assigned to these edges is $4(r-1)$.  In other
words, at a cost of $4(r-1)$, the algorithm has increased the number
of connected components by $r-1$.  Ultimately, when all vertices have
been isolated by edge removals, there are $n$ components; thus, the
total cost of the component creations is at most $4(n-1)$.
\end{proof}

In summary, our estimates ${\tilde k}_e$ satisfy the necessary
conditions for our Compression and Smoothing applications: ${\tilde
k}_e \le k_e$ and $\sum 1/{\tilde k}_e = O(n)$.

\begin{lemma}
  {\tt Estimation} runs in $O(m\log^2 n)$ time on an unweighted graph.
\end{lemma}
\begin{proof}
  Each level of recursion of {\tt Estimation} calls subroutine {\tt
    WeakEdges} on graphs of total size $m$.  An unweighted graph has
    maximum strong connectivity $m$ and therefore has $O(\log m)$
    levels of recursion.
\end{proof}

\subsection{Weighted graphs}

Until now, we have focused on the estimation of edge strengths for
unweighted graphs.  When graphs are weighted, things are more
difficult.  

Nagamochi and Ibaraki give an $O(m+n\log n)$-time weighted-graph
implementation of their {\tt Certificate}
algorithm~\cite{Nagamochi:Mincut}.  (In weighted graphs, the
$k$-sparse sparse certificate has an upper bound of $k(n-1)$ on the
\emph{total weight} of edges incorporated.)  We can use the
Nagamochi-Ibaraki weighted-graph algorithm to implement {\tt
Partition}$(G,k)$ in $O(m\log n)$ time for any value of $k$.  Unlike
the unweighted case, the repeated calls to {\tt Certificate} need not
decrease the \emph{number} of edges substantially (though their total
weight will decrease).  However, the claimed halving in vertices still
happens.  Thus algorithm {\tt Partition} satisfies a recurrence
$T(m,n)=O(m+n\log n)+T(m,n/2)=O(m\log n)$.  Since {\tt Partition} runs
in $O(m\log n)$ time, we deduce that {\tt WeakEdges} runs in
$O(m\log^2 n)$ time.

A bigger problem arises in the iterations of {\tt Estimation}.  In a
weighted graph with maximum edge weight $W$, the $k_e$ values may be
as large as $n^2W$, meaning that $\Omega(\log nW)$ levels of recursion
will apparently be required in {\tt Estimation}.  This can be a
problem if $W$ is superpolynomial.  To deal with this problem, we show
how to localize our computation of strong connectivities to a small
``window'' of relevant connectivity values.

We begin by computing a rough underestimate for the edge strengths.
Construct a maximum spanning tree (MST) for $G$ using the weights
$u_e$.  Let $d_e$ be the minimum weight of an edge on the MST-path
between the endpoints of $e$.  The quantities $d_e$ can be determined
in $O(m)$ time using an MST {\em sensitivity analysis}
algorithm~\cite{Dixon:MST} (practical algorithms run in $O(m\log n)$
time and will not dominate the running time).  Since the MST path
between the endpoints of $e$ forms a (nonmaximal) $d_e$-connected
subgraph containing $e$, we know that $k_e \ge d_e$.  However, if we
remove all edges of weight $d_e$ or greater, then we disconnect the
endpoints of $e$ (this follows from maximum spanning tree
properties~\cite{Tarjan:DataStructures}).  There are at most
$\binom{n}{2}$ such edges, so the weight removed is at most $n^2 d_e$.
Therefore, $k_e \le n^2 d_e$.  This gives us an initial
factor-of-$n^2$ estimate $d_e \le k_e \le n^2 d_e$.

Our plan is to compute the ${\tilde k}_e$ in a series of phases, each
focusing on a set of edges with narrow range of $d_e$ values.  In
particular, we will contract all edges with $d_e$ above some upper
bound, and delete all edges with $d_e$ below some lower bound.  Then
we will use {\tt Estimation} to assign ${\tilde k}_e$ labels to the
edges that remain.

\begin{lemma}
If we contract a set of edges, all of which have weights at least $W$,
then the strengths of edges with original strength less than $W$ are
unchanged.
\end{lemma}
\begin{proof}
Consider an edge $e$ with strength $k_e$, and suppose that its
strength is $k'_e$ in the contracted graph.  It follows that there is
some maximal $k'_e$-connected component $H'$ containing $e$ in the
contracted graph.  Consider the preimage $H$ of this component in
$G$---that is, the set of vertices that get contracted into $H'$.
This component is at best $k_e$-connected in $G$ by the definition of
$k_e$.  It follows that there is some cut of value $k_e$ in this
component.  The edges of this cut have value at most $k_e$, so
contracting edges of value exceeding $k_e$ cannot destroy this cut.
Thus, the connectivity of $H'$ is at most $k_e$.  It follows that
$k'_e \le k_e$.  Since contracting edges cannnot decrease
connectivities, we deduce $k'_e=k_e$.
\end{proof}

We label our edges in a series of phases.  In a phase, let $D$ be the
maximum $d_e$ on any unlabelled edge.  Since $k_e \le n^2 d_e$, the
maximum strength of any unlabelled edge is at most $n^2 D$.  Our goal
in one phase is to (validly) label all edges with $d_e \ge D/n$.  We
begin by contracting all edges of weight exceeding $n^2 D$.  By the
previous lemma, the contractions do not affect strengths of edges with
$k_e \le n^2D$ (which includes all unlabelled edges).  In the
resulting graph, let us delete all edges with $d_e < D/n$ (since $d_e
\le k_e$, no edge we want to label is deleted). The deletions may
decrease certain strengths but not increase them.  It follows that
every unlabelled edge (all of which have $k_e \le Dn^2$) has strength
in the modified graph \emph{no greater} than in $G$.

On each connected component $H$ induced by the remaining edges,
execute {\tt Estimation}$(H,D/n)$.  By \ref{prop:estimation-correct},
this assigns valid lower-bound labels to all edges $e$ with strength
at least $D/n$ (in the modified graph).  In particular, the labels are
valid for all $e$ with $d_e \ge D/n$ (since any edge with $d_e \ge
D/n$ is connected by a path of edges of value at least $D/n$, none of
which get deleted in the phase).  These labels are valid lower bounds
for strengths in the modified graph; however, as discussed in the
previous paragraph, all unlabelled edges have the strengths in the
subgraph no greater than their strength in $G$.  Thus, the computed
labels can be used as valid labels for all the unlabelled edges with
$d_e \ge D/n$.

The approach just described has computed labels for each unlabelled
edge with $d_e \ge D/n$.  We have therefore reduced the maximum $d_e$
on any unlabelled edge by a factor of $n$.  We iterate this process,
continuously decreasing the maximum unlabelled $d(e)$, until all edges
are labelled.

Summarizing our discussion above gives the algorithm {\tt
  WindowEstimation} listed in Figure~\ref{fig:window}.

\begin{figure}
\boxtext{
{\tt {\bf procedure} WindowEstimation}$(G)$

\bigskip

\begin{tabbing}
else \= else \= else \= else \= \kill
\> Sort the edges in decreasing order of $d_e$ into a list $L$\\
\> initialize $G'$ as an empty graph on the vertices of $G$\\
\> {\bf repeat}\\
\> \> {\bf let} $D \leftarrow$ maximum $d_e$ among unlabelled edges in $L$\\
\> \> contract every $e\in G'$ with $d(e)>n^2D$ \\
\> \> move every edge $e \in L $ with $d_e \ge D/n$ to $G'$\\
\> \> call {\tt Estimation}$(G',D/n)$ to get labels ${\tilde k}_e$\\
\> \> \> for the \emph{new} edges added from $L$ in this phase\\
\> {\bf until} no edges remain\\
\end{tabbing}
}
\caption{{\tt WindowEstimation} for weighted graphs}
\label{fig:window}
\end{figure}

\begin{lemma}
Procedure {\tt WindowEstimation} can be implemented to run in
$O(m\log^2 n)$ time.
\end{lemma}
\begin{proof}
The contractions in {\tt WindowEstimation} can be implemented using a
standard union-find data structure~\cite{Cormen:Algorithms}.  Each
time an edge is contracted, a union is called on its endpoints.  Each
time an edge is added from $L$, find operations can identify its
endpoints.  Therefore, the additions and contractions of edges do not
affect the running time.  Instead, the running time is determined by
the repeated calls to {\tt Estimation}.

Consider a particular iteration of the loop with some $D$
value.  We initially contract all edges with $d_e > n^2D$, so that the
maximum strength in the resulting graph is at most $n^4 D$.  We invoke
{\tt Estimation} with a starting strength argument of $D/n$, which
means that it terminates in $O(\log n)$ iterations (the number of
argument doublings from $D/n$ to $n^4 D$).  As to the size of the
problem, recall that we contracted all edges with with $d_e \ge n^2D$
and deleted all edges with $d_e < D/n$.  It follows that our running
time is proportional to $m'\log^3 n$ where $m'$ is the number of edges
with $D/n \le d_e \le D$.

Now we can bound the running time over all phases.  An edge $d(e)$ is
present (neither contracted nor deleted) if and only if $D/n \le D <
n^2D$.  Since the threshold $D$ decreases by a factor of $n$ each
time, this means that edge $e$ contributes to the size of the
evaluated subgraph in at most 3 iterations.  In other words, the sum
of $m'$ values over all iterations of our algorithm is $3m$.  It
follows that the overall running time of these iterations is $O(\sum
m'\log^3 n)=O(m\log^3 n)$.
\end{proof}

\begin{lemma}
Procedure {\tt WindowEstimation} assigns labels such that $\sum
u_e/{\tilde k}_e = O(n)$
\end{lemma}
\begin{proof}
Recall the definition of cost of edge $e$ as $u_e/{\tilde k}_e$.  Our
algorithm incorporates some of the labels computed by {\tt Estimation}
in each phase, contributing their cost (in that phase) to the final
total cost.  We show that the total cost of \emph{all} labels computed
over all the phases is $O(n)$.

We invoke the concept of \emph{rank}.  The rank of a graph is equal to
the number of edges in a spanning tree of the graph.  Inspection of
{\tt Partition} shows that the total weight of edges returned by {\tt
Partition}$(G,k)$ is at most 4 times the rank of $G$.  Similarly,
inspection of {\tt Estimation} show that on a rank-$r$ graph, its
results satisfy $\sum u_e/{\tilde k}_e =O(r)$.

In a phase, we contract all edges of weight exceeding $Dn^2$ and
delete all edges with weight less than $D$.  By the properties of
maximum spanning trees, the resulting graph is precisely spanned by
the set of MST edges with weights in this range.  That is, the rank of
this graph is equal to the number $r_D$ of such MST edges.  It follows
that the total cost $\sum u_e/{\tilde k}_e$ of {\tt Estimation} labels
in this phase is $O(r_D)$.  Now note that each MST edge contributes to
$r_D$ only when its weight is between $D$ and $Dn^2$, which happens in
at most 3 phases since $D$ decreases by $n$ each phase.  Thus, each
edge contributes to 3 $r_D$ values, so $\sum r_D \le 3(n-1)$.  This
bounds the total cost by $O(\sum r_D)=O(n)$, as desired.
\end{proof}

\section{Conclusion}

We have given new, stronger applications of random sampling to
problems involving cuts in graphs.  The natural open question is
whether these approximation algorithms can be made exact.  An initial
step towards the answer was given in~\cite{Karger:Skeleton}, but it
only gives a useful speedup for graphs with large minimum cuts.  More
recently, sampling has led to an exact linear-time algorithm for
minimum cuts~\cite{Karger:Lincut}; however, the techniques used there
appear to be specialized to that particular problem.  Karger and
Levine~\cite{Karger:AugmentingPath} have recently given very fast
algorithms for flows in unweighted graphs; the important remaining
question is to develop fast exact algorithms for weighted graphs.

A more limited open question has to do with the use of strong
connectivity.  We introduced strong connectivity in order to make our
theorems work.  Many of the intuitions about our theorems apply even
to the standard connectivity notion in which the connectivity of edge
$(u,v)$ is defined to be the minimum $u$-$v$ cut in $G$.  We have no
counterexample to the conjecture that using these weak connectivities
would suffice in our algorithms.  Such a change would likely simplify
our algorithms and presentation (though the time bounds are unlikely
to change).

\appendix

\section{The General Weighted Sampling Theorem}

For possible future use, we give a general theorem on when a weighted
random graph has all cut values tightly concentrated near their
expectation.  The compression theorem and smooth graph sampling
theorems are special cases of this theorem.

\begin{theorem}
  Let $G$ be a random graph in which the weight $U_e$ of edge $e$ has
  a probability distribution with expectation $u_e$ and maximum value
  $m_e$.  Let $k_e$ be the strength of edge $e$ in the graph where
  each edge $e$ gets weight $E[U_e]$.  If for every edge, $k_e \ge
  2m_e(\ln n)/\epsilon^2$, then with high probability, every cut in
  $G$ has value within $(1\pm\epsilon)$ times its expectation.
\end{theorem}
\begin{proof}
  Order the distinct edge strengths $k_1,\ldots,k_r$ in $H$ in
  increasing order.  Let $F_i$ be the graph consisting of all of
  $k_i$-strong edges in $H$, with edge $e$ given weight $U_e/k_e$ (so
  $F_i$ is a random graph).  Observe that $G=\sum(k_i-k_{i-1})F_i$.
  So if every $F_i$ is near its expectation, it follows that $G$ is
  near its expectation.

  So consider (a component of) graph $F_i$.  The expected value of a
  cut in $F_i$ has the form $\sum u_e/k_e \ge 1$ by
  \ref{prop:convergent-sum}.  In other words, the minimum cut in
  $E[F_i]$ is at least 1.  On the other hand, the maximum value
  attained by any edge in $F_i$ is $U_e/k_e \le m_e/k_e$.  By
  \ref{prop:basic-sampling}, it follows that $F_i$ has all cuts within
  $(1\pm\epsilon)$ of its expectation with high probability.
\end{proof}

\bibliographystyle{alpha} \bibliography{me}

\end{document}